# Networks of digital humanities scholars: The informational and social uses and gratifications of Twitter




## Abstract

Big data research is currently split on whether and to what extent Twitter can be characterised as an informational or social network. We contribute to this line of inquiry through an investigation of digital humanities scholars' uses and gratifications of Twitter. We conducted a thematic analysis of 25 semi-structured interview transcripts to learn about these scholars' professional use of Twitter. Our findings show that Twitter is considered a critical tool for informal communication within DH invisible colleges, functioning at varying levels as both an informational network (learning to 'Twitter' and maintaining awareness) and a social network (imagining audiences and engaging other digital humanists). We find that Twitter follow relationships reflect common academic interests and are closely tied to scholars' pre-existing social ties and conference or event co-attendance. The concept of the invisible college continues to be relevant but requires revisiting. The invisible college formed on Twitter is messy, consisting of overlapping social contexts (professional, personal, and public), scholars with different habits of engagement, and both formal and informal ties. Our research illustrates the value of using multiple methods to explore the complex questions arising from big data studies and points toward future research that could implement big data techniques on a small scale, focusing on sub-topics or emerging fields, to expose the nature of scholars' invisible colleges made visible on Twitter.


## Keywords

Uses and gratifications; digital humanities (DH); scholars; Twitter; social media; big data; social networks

## Introduction

Communication is an integral part of scholarly practice. Indeed much work in the sociology of science has stressed that disciplines only flourish if an invisible college is established to facilitate communication and exchange among its members. De Solla Price coined the term "invisible colleges" in 1961, which describes "a communication network of a subgroup of researchers within a research area" (Crane, 1972, p. 35). While citation networks are one way to identify and study an invisible college, much research advocates for a stronger focus on informal ties that sustain disciplinary discourse, conversation, and collaboration (Carroll & Dasler, 2015; Lievrouw, 1990; Zuccala, 2006). With the widespread popularity of social media in academia, it is relevant to examine the role of informal ties to test out new ideas, circulate pre-publications for constructive feedback, and discuss emerging topics (Gruzd, Staves, & Wilk, 2012; Holmberg & Thelwall, 2014; Rasmussen Neal, 2012; Veletsianos, 2012). An important by-product of the flurry of social media activity is the cloud of big data it produces and the user interactions revealed. Researchers have begun to take advantage of big data approaches to uncover patterns and topological features of networks. We argue that it is also important to examine the social context in which these patterns emerge because it can inform and expand findings from big data analytics.

Through methods such as interviews and surveys, prior research has examined scholarly use of social media (Gruzd, Staves, & Wilk, 2012; Nicholas & Rowlands, 2011), but few studies have explicitly explored the use of Twitter, a popular microblogging platform, by scholars (Holmberg & Thelwall, 2014). Given the popularity of Twitter among scholars in various disciplines (Holmberg & Thelwall, 2014; Nicholas & Rowlands, 2011) and the importance of scholarly communication to individual scholars and to communities of researchers (Gruzd, Wellman & Takhteyev, 2011; Holmberg & Thelwall, 2014), it is

critical to understand how and why Twitter is being used and its implications in relation to scholarship. Because of the dynamic nature of social media (see Hogan & Quan-Haase, 2010), research also needs to consider that social media is not only used differently across platforms (Puschmann & Bastos, 2015; Archambault & Grudin, 2012), but also across disciplines (Holmberg & Thelwall, 2014). Much of the literature has tended to look at higher education (Carroll & Dasler, 2015; Veletsianos, 2012) or at scholars as general groupings (Nicholas & Rowlands, 2011), without adequately investigating the social ties that connect scholars and make their interactions socially relevant nor recognizing potential differences in uptake among disciplines (Holmberg & Thelwall, 2014). For example, digital humanities (DH) and cognitive science scholars tend to use Twitter more for conversation than those in disciplines such as biochemistry and economics (Holmberg & Thelwall, 2014). Current discussions on social media use in academia would thus benefit from studies that investigate the adoption and usage of specific platforms by subgroups of researchers.

The use of Twitter is still relatively new, and thus researchers are only beginning to grasp Twitter's "antecedents and consequences" (Aladwani, 2015, p. 15). One central debate in the literature, informed in part by big data research (Kwak et al. 2010; Myers et al., 2014), has emerged around how Twitter exchanges provide informational or social benefits to users. Kwak et al. (2010) suggest that Twitter is best described as an information network rather than a social network. More recent work examining the topological features of networks found evidence that suggests Twitter acts as both a information and social network (Myers et al., 2014). To discern the informational and social gratifications obtained from Twitter networks, it is vital to study Twitter activity by users who have shared interests and are bound by informal ties. The aim of this paper is to inform and expand findings based on big data analyses about Twitter's informational and social gratifications.

Our research contributes to the growing body of literature on scholars' use of social media by examining a single group of scholars, those in the DH community who use "computational tools to do the work of

the humanities" (Unsworth, as quoted in Gold, 2012, p. 67). Our findings inform researchers, educators, and policy makers who strive to understand the use and value of social media in the context of scholarship (e.g., Gruzd et al., 2012; Holmberg & Thelwall, 2014). The current study examines two research questions:

(a) How are DH scholars using Twitter in the context of their scholarly practice?

(b) What informational and social gratifications do DH scholars obtain from their usage of Twitter?

Our two research questions emerge from prior research in the area of big data analytics. We specifically examine whether, in the context of DH, Twitter's uses and gratifications are more informational or social in nature. Big data analytics are particularly well-suited for identifying the topological features of a network, but are myopic to the social context in which interactions take place. The contribution of this paper is to demonstrate how a small scale study of Twitter uses and gratifications can inform and expand a debate unfolding in big data analytics. Thereby, this paper sheds light on the debate by taking a look at a small slice of the activity on Twitter from the perspective of DH scholars.

## Background

*Big data and Twitter use*

Though criticised for its biases and assumptions, big data informs our understanding of social phenomena and, at the same time, creates "a radical shift in how we think about research" (boyd & Crawford, 2012, p. 665). Even though the definition of big data is contentious and evolving, the term is generally used to describe large and complex data sets, including content generated on social media sites (Big Data & Society, n.d.). Big data studies have compared a range of topological features of Twitter networks; a study by Kwak et al. (2010) being "the first to look at the entire Twittersphere" (p. 599) for the purpose of

investigating user participation in trending topics, reciprocity, and homophily. Kwak et al.'s study demonstrates how the use of big data techniques can provide new insights into large-scale social processes. A key finding was that the majority of ties on Twitter were one-way rather than reciprocal. Based on their study, the authors proposed that Twitter's primary use is as a source of information and platform for information exchange rather than social networking. In another study, Myers et al. (2014) found that while some of their results pointed to Twitter as a social network, others were more consistent with Twitter as an information network, thus partially corroborating Kwak et al.'s (2010) findings, but ultimately concluding Twitter represents more of a *hybrid* network. Myers et al. (2014) further hypothesized that Twitter starts off as an information network but over time evolves into a social network as a user becomes more selective of who to follow: "The user typically discovers a community with which to engage—whether it be based on real-world social ties, common interests, or other factors—and Twitter starts behaving more like a social network" (p. 498).

Big data studies (e.g., Kwak et al., 2010; Myers et al., 2014) provide useful generalizations about patterns in human behaviour on Twitter. However, subsets of Twitter users may not exhibit the same patterns and these nuances may be lost in analyses of large data sets. Holmberg and Thelwall (2014) selected researchers from ten different disciplines, including DH, biochemistry, and economics, based on their productivity and presence on Twitter. Based on an analysis of 200 randomly selected tweets from these researchers, they found significant differences among disciplines in terms of how links were shared, the number of conversational tweets, retweets, and tweets. Better understanding the role of social context in how different social groups make use of Twitter and the unique gratifications each group gains from their engagement will help researchers interpret the 'big picture' findings drawn from big data analytics.

*The digital humanities (DH) community*

DH is an interdisciplinary, evolving field of study with roots in the humanities computing centres of the 1970s and 1980s (Hockey, 2004). Developments in computing have allowed for a much larger integration of technology into the humanities, making possible the collection, analysis, and interpretation of big data that can span several centuries, time periods, and data sources. The central tenet of DH is the utilization of computing resources as a means to aid scholarship, teaching, and theorizing around humanities subjects (e.g., language, literature, history, art, and philosophy) and there is a growing, international network of scholars who identify themselves as members of the DH community (Kirschenbaum, 2010).

*DH and social media*

Definitions of social media abound and tend to include a wide range of tools such as micro-blogging, social networking sites, and video sharing websites (Hogan & Quan-Haase, 2010). What characterizes social media engagement is that it is global in nature, links loosely connected networks, facilitates the sharing of ideas and content in an interactive manner, and allows for high levels of engagement with people and content (McCay-Peet & Quan-Haase, forthcoming).

Not all scholars have adopted social media, but for many it has become an integral component of their scholarly practice (Quan-Haase, 2012). Social media reduces the barriers of time and location to scholarly communication; scholars can tweet messages to their followers and have answers to questions in real time; they can communicate with a co-worker who is away at a conference; or they can follow activities of a conference they were unable to attend in person from afar (Holmberg & Thelwall, 2014; Ross et al., 2011). DH scholars have been described as early adopters and avid users of social media (Kirschenbaum, 2010; Ross et al., 2011) and many DH scholars embed Twitter in their practice to amplify scholarly communication (Howard, 2009). Schnapp and Presner (2009) described DH as "iterative scholarship, mobilized collaborations, and networks of research," (n.p.) which is illustrated through DH scholars' use

of social media to communicate with their colleagues and share research results with a broad audience (Terras, 2012). Twitter, for example, plays a large role during DH events, workshops, THATCamps, and conferences as a platform for backchanneling (Ross et al., 2011). Through a content analysis of tweets by scholars from ten disciplines, Holmberg and Thelwall (2014) found that DH scholars tend to engage in more conversations (@username), share fewer links in their tweets, and use Twitter for scholarly communication more so than scholars in some other disciplines. Based on a sample of 200 DH scholar tweets, the study also found that the ten most frequently-used words in tweets by DH scholars included: "need," "thanks," and "thinking" as well as "digital" and "humanities," which suggests that DH scholars are sharing ideas and information related to DH in an interactive manner.

In summary, scattered throughout the literature are indications that Twitter is an important tool for DH scholars. What is a more holistic understanding of DH scholars' motivations behind Twitter use (e.g., informational versus social), how they are using Twitter as a platform for scholarly communication, how they are navigating the use of Twitter in the context of their professional work, and to what effect.

## Methodology

This study employed semi-structured interviews to collect data from 25 DH scholars about their use of digital tools and social media for supporting their scholarly work. Transcripts were analysed using the thematic analysis technique and the uses and gratifications framework informed the coding process.

### Participants

Because debate surrounds what DH entails as well as the creation of arbitrary boundaries within the humanities, participant sampling was grounded in practical criteria that would allow us to recruit widely and be inclusive. We were also concerned about recruiting individuals with various DH experiences (e.g., faculty, graduate students, and librarians). For this purpose, we recruited at the international DH annual conference (DH 2013), which took place in Lincoln, Nebraska, July 16 to 19, 2013 ($n = 15$) and through

site visits to two DH centres in the United Kingdom and Ireland in February 2013 ($n = 5$) and June 2013 ($n = 5$), respectively. Participants were recruited through four strategies: a) in person, b) Twitter, c) email prior to site visits, and d) a booth set up at the DH 2013 conference.

Participants ($N = 25$) comprised an international set of scholars engaging in DH-related scholarship and included 14 females and 11 males. Eleven professors, six PhD students, three managers, two librarians, two public historians, and a retired professor participated. The majority had attained a graduate degree: 11 had a PhD and 11 had a MA degree, and three participants had attained an undergraduate degree. Participants' fields of study varied and reflected the diversity existent within DH scholarship: classics, literature, English, history, art history, information science, computer science, engineering, political communication, and business. Participants were from the U.S., Canada, Germany, England, Ireland, and Australia.

*Procedure*

Each in-person, 45 to 60 minute session followed the same procedure beginning with an introduction to the study and consent. The interviews were audio-recorded for later transcription. Interviews were semi-structured to allow the interviewer the flexibility to clarify participant responses, probe for more detail, and follow-up on unexpected topics (Berg, 2007). The interview was part of a broader study on the information behaviours of digital humanists. The present analysis is limited to the sections on Twitter and general social media use (see Appendix).

*The uses and gratifications framework*

The uses and gratifications approach was selected as a framework for analysis. Uses and gratifications is "a psychological perspective that examines how individuals use mass media, on the assumption that individuals select media and content to fulfill felt needs or wants" (Papacharissi & Mendelson, 2011) and

is an ideal method for the study of social media in the context of everyday life (Quan-Haase & Young, 2010). The uses and gratifications approach has been used to identify predictors of time spent on social media such as 1) relaxing entertainment, 2) expressive information sharing, and 3) social interaction (Smock et al., 2011). In a similar vein, the uses and gratifications approach was used to examine the perceived characteristics of Twitter as a motivating factor for its continued use (Coursaris et al., 2013). The approach is also useful for making cross-platform comparisons. A comparison of the uses and gratifications of Facebook and instant messaging revealed several reasons why young people integrate multiple digital technologies into their daily lives (Quan-Haase & Young, 2010). While a significant portion of prior research on social media has been grounded in the uses and gratifications approach, we know of no prior study that has applied the uses and gratifications approach in the context of scholars' use of Twitter.

*Data analysis*

We performed thematic analysis of the transcripts (Braun & Clarke, 2006), involving inductive and deductive coding. Thematic analysis allowed the researchers to use the uses and gratifications framework to help guide the analysis and draw on prior research to identify pre-existing themes while at the same time providing the flexibility to develop new themes. Thematic analysis is an iterative process that involves familiarization with the data (via transcription, reading of transcripts), generation of initial codes, a search for themes, review of themes, and the naming and definition of themes. Data analysis was guided by the two research questions and thus carefully examined participant references to the uses and gratifications of Twitter. To help ensure rigour in the data analysis process, multiple coding was used (Barbour, 2001). Multiple coding involved the analysis of the transcripts by the second and third authors. The authors completed the analysis separately while discussing codes together after every five or so transcripts to help check for agreement and, more importantly, to discuss insights and alternative interpretations. Fifteen of the twenty-five transcripts were coded using this method. A further 10

transcripts were analysed using the codes and no further major themes were identified, thus reaching

saturation. Descriptive statistics drawn from the Twitter use set of questions (e.g., whether or not Twitter

is used to connect with people within the DH community) were relied upon to complement and

corroborate the themes identified in this research. Following recommendations outlined in Braun and

Clarke (2006), writing-up the results section provided a final opportunity for analysis, as representative

quotes from participants were extracted from the transcripts to highlight key findings.

**Findings**

*Social media and Twitter use*

Participants reported varying levels of use of social media. All 25 participants maintained at least one

social media account (e.g., Twitter, Facebook, Academia.edu), but many maintained several. All but one

participant had a Twitter account and one participant had a Twitter account but was not currently using it.

When asked about their Twitter networks, 19 participants recalled their number of followers, which

ranged from five to 2,000 (m = 550; sd = 667) and 16 recalled the number of people they followed, which

ranged from 8 to 2,000 (m = 510, sd = 670) and 12 of these 16 participants reported following more than

150. One hundred and fifty is generally considered the maximum number of stable social relationships

that can be maintained at one time (Dunbar, 1992, as cited in Myers et al., 2014). This suggests that DH

Twitter ties are not entirely social in nature for the majority of DH scholars. While we did not collect data

on the reciprocity of the ties, ten reported following more people or organizations than they were

followed by, five followed fewer people or organizations than they were followed by, and one had an

equal number of followers and follows.

Several participants reported their time investment in Twitter in relative time spent (e.g., "not a lot,"

"probably too much"), others spoke of the quality of their investment (e.g., "interstitial," "constant," "a

lot, but uneven"), and yet others reported on the context of usage referring to "only when not busy with

work" or "not until recently at the conference." Seven participants reported tweeting on a daily basis, five

reported tweeting at least once per week, while three reported that it varies and one reported rarely tweeting. In other words, there was no single pattern of Twitter usage among participants; they adapted the technology to their own social and professional needs.

## Twitter's informational uses and gratifications

In order to share information with or solicit information or feedback from those in their Twitter networks, DH scholars develop skills to do so effectively (Learning to Twitter). Likewise, to reap the benefits of a rich information network, DH scholars use Twitter to keep up with news and information in the DH community (Maintaining awareness).

### Learning to 'Twitter'

In a recent article on 'Twitteracy', Greenhow and Gleason (2012) described the multimodality of microblogging: images, links, emoticons, and standardized language comprise the 140-character tweets users compose. The authors also noted that there are a number of conventions (e.g., hashtags and retweets) that users need to know to effectively participate on Twitter. Our findings similarly suggest that learning how to engage in a multimodal digital environment is central to participants. P19 noted the importance of selecting the "right" hashtag for a tweet to maximize its reach. We found that simply learning to write in the allocated 140-character limit was critical for meaningful engagement.

The 140-character limit was seen as beneficial to some, forcing them to think through their ideas more clearly and to express themselves succinctly, but the character limit was too restrictive for others. In most cases, our participants weighed both sides of this "new literary practice" (Greenhow & Gleason, 2012): they recognized the challenges, but were still willing to engage. Participants reported that the effort of "crafting tweets" (P6) was worth it because they were learning a new digital literacy. Several referred to writing posts as a *skill* that could be taught and improved through experience.

Gruzd et al. (2011) refer to 'twitterspeak' as the linguistic conventions utilized to tweet. Participants in the present study perceived twitterspeak as a new means of communicating that encouraged them to think through their tweets.

> I think there is a literacy here because it forces you to think about being economical with your mode of expression. What is the word that matters most? What happens if I move this? Is that going to, you know, impact it in important ways? You can get the sense across, but do you get the nuance? [...] What are you losing? (P6)

The word limit was also perceived as an opportunity to 'boil down' ideas to their essence. Eleven participants indicated that they use Twitter to help them clarify ideas, while two indicated that they use it for this purpose 'sometimes' or 'somewhat'. The need to concisely express complex ideas seemed a welcome challenge to some participants, one that allowed them to clarify their thinking and reflect upon their ideas. This suggests that DH users of Twitter optimize the informational value of the platform and realize its potential for the exchange of information.

Some participants stressed the shortcomings of the medium by stating how it is difficult to write within the 140-character format, how it stifles conversation, and how it leads to heavy reliance on links for additional information. In that regard, they saw their tweets interconnected with larger discourses occurring on blogs, articles, and other web content. Twelve participants described the constraints imposed by the 140-character limit in expressing complex ideas and conveying large amounts of information.

> I mean obviously a conversation is better face-to-face. Sometimes it's hard for me to say what I want to say in 140-character snippets. Especially if there are a lot to say [sic]. (P3)

Although several participants noted that there was an interactive, conversational element to being on Twitter, several agreed with P3 that conversations were best left for other media. This highlights the tensions in relying on Twitter for the purpose of sustaining an imagined community: Twitter facilitates the easy and convenient sharing of information, but it is also important for the messages to make sense. Hence, to be a part of this imagined community, members must first share a common language (Anderson, 1983) and set of conventions.

In sum, while the majority of our participants have adopted Twitter, they continue to figure out how it best fits into their scholarly practice. They are willing to further push themselves to understand the intricacies of the medium and to learn how to engage with others in creative and meaningful ways that will allow them to take advantage of Twitter as an information network.

*Maintaining awareness*

Prior research (Coursaris et al., 2013; Dantonio et al., 2012; Kwak et al., 2010) underlines the importance of Twitter for helping people *maintain awareness* of news and information, a dominant theme identified in our data set as well. Many participants noted the importance of maintaining awareness of DH-related research, conferences, and projects via Twitter. One participant asserted Twitter was unrivalled in this respect and this confirms prior research that stresses the centrality of Twitter as an information conduit (Kwak et al., 2010).

> I would say that Twitter is probably the single most important place for me to keep up with new websites, new projects (P1).

But more than the *passive* type of information behaviour often described in the literature (Wilson, 1999) and that the concept of 'maintaining awareness' perhaps suggests, one participant (P5) observed how

useful Twitter was for quickly getting up to speed on a particular topic by employing a range of *purposeful* information tactics.

> When I first started on the [*DH Project*], I needed to know more, I needed to 'skill up' very quickly. So I actively followed people that I thought might be helpful for extending my knowledge in that area.

Twitter was more than just background noise. It presented itself as an information space that could be mined for *purposeful* information seeking. The DH information that could be obtained from Twitter clearly motivated participants' continued use of the site and as new information needs emerged, participants utilized Twitter strategically to fill their knowledge gaps.

Three further interrelated sub-themes of maintaining awareness were identified: time, noise, and serendipity. These three sub-themes point to motivations for continued Twitter use as well as potential reasons for avoiding Twitter, at least at times (Makri et al., 2014).

Maintaining awareness requires an investment of time and for some participants it was difficult to find the time during a regular workday to check their accounts and thus reported using Twitter primarily during conferences or events (see Engagement section). Others reported how easy it was to maintain awareness of DH information by periodically checking their Twitter timelines. One participant noted how easy it was to just "slot it in" (P25) and another mentioned using "it in those in-between moments when I am waiting for things or walking down the hall on my way between meetings" [P23]. The way some participants described how Twitter helped them maintain awareness had a *ubiquitous* quality to it.

> Well for me, I have two screens. One of my screens is purely devoted to Twitter. It is on all talk, all the time. [...] I use it every single day [...] all day. People say that's incredibly difficult and

people don't really enjoy having it there constantly, but I find it a brilliant help [...] not only in

procrastination, but in hooking into things that I don't necessarily know about. (P21)

P21's observations about her own use of Twitter together with her perception of how others would react

to having Twitter in the background at all times illustrates two sides of the same coin. The negative side is

the distraction resulting from all the noise—the massive amounts of diverse, but not always interesting or

useful information that has a tendency to overload users' timelines. This dissuaded some of our

participants from being continuously connected to Twitter. One participant noted that what he did not like

about Twitter was the inconsistent quality of content on his timeline, which was often filled with

uninteresting, irrelevant information.

Ridiculously long list, with ridiculously useless information and that's the danger of combining

the personal account with the professional. [...] Research ones are much more focused but I find

that people just tweet all the time and sometimes I don't need to read that. (P7)

Noise also relates to the theme of the 'imagined audience' (Marwick & boyd, 2010) explored in the

section to follow. People may imagine their audience and speculate what posts people in their network

would find interesting, but it is difficult to optimize information for a diverse network. There is an

inevitable disconnect, at least periodically, between the imagined audience and the actual audience.

Finally, the positive side of this diverse and dynamic digital environment is the serendipitous, unexpected

experiences it affords. Ten participants reported that they use Twitter for making random and important

connections. Two participants specifically referred to the serendipitous quality of their experiences. P5

reported dealing with a problematic aspect of her research and 'stumbling upon' a tweet directly related to

the problem.

In summary, Twitter is an important, however imperfect, information network to our participants. Informal communication via Twitter—at times useful and others useless—helps participants maintain awareness of developments in the field of DH. Next we discuss how some of our participants not only use Twitter as a source of information, but as a social network to engage with fellow DH scholars.

## Twitter's social uses and gratifications

Complementing the uses and gratifications of Twitter for DH scholars as a critical, yet informal information network was the use of Twitter as a social network. Participants were keenly aware that others were reading their tweets (Imagined Audiences) and acted accordingly. Moreover, many participants also recognised the value of using Twitter's affordances to engage on a personal level with their social network (Engaging other Digital Humanists).

### Imagining audiences

Participants' adoption and continued use of Twitter appears motivated by the flexibility Twitter affords for creating multiple profiles to reflect different aspects of a scholar's life. An important role played by Twitter was to help participants compartmentalize their various identities, social roles, and social networks through the creation of multiple accounts, which could cater to different audiences. The concept of the 'imagined audience' was coined to describe the difficulties inherent in predicting who will read a specific tweet and thus determining who comprises one's audience (Marwick & boyd, 2010). For Marwick and boyd, "[t]his audience is often imagined and constructed by an individual in order to present themselves appropriately, based on technological affordances and immediate social context" (p. 115). One participant reported having different goals, vocabulary, and subject matter to present to each of the audiences of his two Twitter accounts. He consciously crafted tweets (see Learning to Twitter section) and decided what to retweet based on who he imagined would read these.

I have two Twitter accounts. So I have myself as myself [@*myself*]. But then I also just decided when I became associate dean, I decided to create a dean account @dean*Twitter account*. [...] I wanted a more official voice, so on my @*myself* [...], I blur the boundary between my personal and professional identity (P6).

Though it is difficult to deduce from the findings of this study whether single or multiple Twitter accounts are the norm amongst DH scholars, even those who maintained a single account were aware of who their potential audience could be. P24's maintenance of a single account appeared to be for both a professional and personal audience.

I kind of 'flag-up' whenever I see cool things by/about computing and be like 'oh that's the stuff I retweet most of all,' so [...] if anyone is following me that doesn't know much about digital humanities, they might get a flavour for it (P24).

P24 appears to reconcile posting information that may not be relevant to her entire imagined audience with the belief that those in her personal network, who most likely are unfamiliar with DH, may find it interesting to learn more about it. This notion of posting information deemed relevant by an imagined audience is supported by our findings: most of our participants reported finding information useful to their research ($n = 20$) or information they were otherwise unaware of ($n = 23$). The informational benefits of Twitter are closely linked to the crafting of an "imagined audience" that shares common interests and has overlapping professional and social roles.

Imagined audiences are an important part of the development of a DH scholar's voice on Twitter, but the imagined audience changes over time as users get a sense of who is reading their tweets and how the audience responds to them. By taking notice of what content is retweeted, favourited, or sparks conversation, users gain a better sense of who their audience is and can adapt the content of their tweets

accordingly. Gaining a sense of one's audience over time was evident when P5 spoke of the different results obtained from crowdsourcing using Facebook versus Twitter:

> I might throw a question out to both Twitter and Facebook and you will often get very different answers in those two communities. And so in that sense it can be quite complementary. But they are also therefore quite separate because they are actually producing different things. (P5)

In this sense, imagined audiences can become more visible and tangible over time as a result of feedback obtained through interaction and engagement.

*Engaging Other Digital Humanists*

Following or participating in a Twitter conversation related to a conference is termed backchannel and is a common practice among academics (Knight & Kaye, 2014; Reinhardt, Ebner, Beham, & Costa, 2009). It is most often done by following a hashtag related to the event (e.g., #Thatcamp) and Ross et al. (2011) noted that it is a central motivation for DH scholars to use Twitter. Backchannel during conferences among the DH community was a central sub-theme that addressed the gratifications obtained from engaging with others on Twitter. In the interviews that took place during the DH 2013 conference, participants noted the use of social media throughout the event, even going as far as adopting separate social media management tools, such as Tweetdeck or HootSuite, to be able to follow multiple, simultaneous Twitter conversations.

> I use it mainly at conferences, so I will break out the TweetDeck for the conferences, and then the rest of my time I just check in on my phone maybe once a day. (P8)

Backchannel also allows those individuals who are not physically present at the event to follow the conversation and engage at a distance. Our research supports these findings, as conferences were one of

the main times that our participants indicated relying on Twitter for engagement, and P14, for example, had decided to join Twitter precisely to participate in these online discussions. Reinhardt et al. (2009) found that Twitter was used at conferences mainly for sharing resources and connecting with others, and also for following parallel sessions. Our work supports this as well, as P12 notes:

> At the conferences they always exchange a URL or something. Yeah, that is really useful.

Communication and resource sharing via Twitter are forms of engagement that, for our participants, continue to be important beyond the use around a specific event. Twenty participants reported using Twitter for connecting with other DH scholars on a regular basis.

> It's like the de facto [...] I don't know many people in DH that aren't on Twitter. (P3)

As far as uses for Twitter within this community, all 23 of those participants who reported using Twitter regularly claimed to use it for sharing other's research, which is easy to understand given the ease of the one-click "retweet" button. Eighteen participants chose to engage with their audience by using Twitter to disseminate their own research, a practice that correlates with increased readership (Terras, 2012). More informal exchanges also appear facilitated by Twitter. As previously mentioned, Twitter was used by DH scholars to seek input and advice. One participant (P19) commented on the usefulness of relying on hashtags to gain the attention of specific communities active on Twitter such as #twitterstorians.

Twitter use served to expand participants' social networks by meeting new DH scholars and gaining familiarity with their research interests. Seventeen of the 23 participants who use Twitter claimed they had met new people on Twitter, with P23 even claiming,

> There are some people who I would count as my friends that I've never met in real life.

What is it about Twitter that promotes sharing research, asking questions, and delving into conversations with strangers? Perhaps, as Marwick and boyd (2010) note, it is the shift from a broadcast audience, a "faceless mass," to a networked audience, which is "unidentified but contains familiar faces; it is both potentially public and personal" (p. 129).

When DH scholars engage with one another on Twitter, they are able to acquire more information about a specific user by looking at their website or blog (usually linked to in their Twitter profile), examining other conversations they have had, and finding images or pictures they have posted. This contextualization moves the audience from one that is wholly 'imagined' to one that is identifiable and easier to connect with. Also central to understanding the social gratifications of Twitter is the social context in which scholars interact: DH scholars share similar interests, are part of overlapping networks, use a common language, and have shared goals. This social context allows for the development of a loosely connected social network integrates both strong and weak ties.

## Discussion and Future Work

Our first research question examined how DH scholars use Twitter. Twitter was often part of a broader 'social media toolbox' that includes other social media sites such as Academia.edu (Quan-Haase & Young, 2010). Twitter use varied in the study population, some scholars did not use it a lot, while others felt that they used it too much. DH scholars reported great variability in use: for some use was interstitial, while others said it was a daily activity and yet others said that they used it a lot, but in an uneven manner. Our findings show that sheer numbers cannot reflect how DH scholars engage on Twitter, scholars' thoughts on their use of Twitter were always in relation to how Twitter fits with their current personal and work context. Future research could develop a typology of DH Twitter users, which could serve to make comparisons across disciplinary fields.

The second research question investigated what informational and social gratifications DH scholars obtained from Twitter. As Myers et al. (2014) suggest Twitter is a hybrid network, exhibiting markers indicative of both an informational and social network. We found that the uses and gratifications framework was a fruitful approach for shedding light on this debate. The gratifications obtained range from the more informational gratifications of digital literacy and staying up-to-date on what is happening in DH to the more social gratifications of finding a voice and becoming an active member of the DH community, corroborating and expanding on Myers et al.'s (2014) original work.

Learning to twitter was perceived by DH scholars as a form of digital literacy on which many of them thrived. Often seeing the need to craft tweets as a challenge, their ability to communicate via this platform and make their posts clear to others meant they were able to relay information and ideas and get their main points across. As DH has been shown to be one of the disciplines that relies on Twitter most (Holmberg & Thelwall, 2014), the need to gain this digital literacy skill is critical for being a member of the community. Also of importance is the need to be able to read and interpret tweets in order to maintain awareness about the intricacies of DH. Participants noted that Twitter was the first place they went to when wanting to learn more about DH. Having this ever-changing, constant feed of updates about their DH colleagues, teaching, and research was one of the main informational gratifications these scholars obtained from using Twitter. However, maintaining awareness inevitably leads to exposure to the noise of all of the information those in a person's network deem interesting to share with their 'imagined audience.' Should we develop more sophisticated tools for filtering the noise, to narrow the gap between the imagined and real audience? Perhaps filtering tweets so that only those tweets directly related to a scholar's research interests appear in the scholar's Twitter feed. Or is the noise a double-sided coin, both a strength and a weakness of the social media platform? While too much noise can be bad, noise is also a sign of a dynamic, information-rich environment (Makri et al., 2014). Our findings suggest noise has the potential to facilitate serendipitous experiences.

Striving to communicate more effectively on Twitter also points to ways in which DH scholars are trying to enhance the *social* part of the information/social network equation. There is incentive for DH scholars to join Twitter, evidenced from its current widespread adoption.

What makes communication on Twitter challenging for many of our DH scholars is that it is not only geared toward exchanging information effectively, but also toward engaging in deeper conversations, reaching out to others, providing and giving feedback, and acknowledging support from followers.

Twitter for many DH scholars is a conversation, one that takes place through a single tweet or a series of interactive tweets, and in relation to other discourses online and offline both inside and outside their disciplines. Conversations, retweets, and follows on Twitter all help DH scholars to feel they are part of the community, which keeps its strength through digital media. Future research could examine whether Twitter generates new collaborations and innovative research outcomes, key ingredients for a striving invisible college. Twitter adds to other means of interaction, enhancing through digital connections ties that are formed in other social spheres.

## Conclusion

Welles (2014) made the case for examining subsets of big data sets. She argued that while big data enables the examination of human behaviour at unprecedented scale, it also provides an opportunity to examine minorities and outliers. While big data enables researchers to summarize human behaviour, researchers should not lose sight of a second major advantage of big data, "the plentiful representation of minorities" (Welles, 2014, p. 1). Though it remains to be seen whether DH scholars are outliers with respect to their behaviour on Twitter, big data analysis techniques could provide an opportunity to increase our understanding of the inner workings of the DH invisible college within Twitter.

We conclude that the concept of the invisible college as introduced by de Solla Price requires revisiting. The invisible college formed on Twitter is messy, consisting of overlapping social contexts (professional, personal, and public), scholars with different habits of engagement, and both formal and informal ties.

Future research could investigate how the invisible college formed on Twitter is associated with citation and co-citation networks (Mutschke & Quan-Haase, 2001). We also conclude that our findings support Myer et al.'s (2014) notion of a hybrid network (social and informational). Furthermore, our study informs and expands big data research and illustrates that social context matters. Through its macro-level analysis, big data has framed Twitter networks as a dichotomy—information versus social network. Our research provides a more nuanced understanding of how shared interests (e.g., common subject specialties), goals, understandings, and pre-existing social ties can influence the nature of the network. Within the context of DH scholarship, information networks do not function solely as a means for broadcasting to and gathering from a large information network of followers. Rather, it seems that ties that may start as information-based ties often become social ties in the context of finding shared ground.

**Acknowledgement:** We would like to thank two anonymous reviewers and the editors of *BD&S* for the insightful feedback. This project was funded by a Social Science and Humanities Research Council of Canada [grant number R3603A13].

**Appendix Interview Guide**

*Note:* All six main sections of the interview guide are included below, but only those questions that relate to Twitter usage are outlined in full.

1. *Opening Questions* [Academic status, education, current position]

2. *The Digital Humanities Centre* [DH centre or project with which participant is affiliated]

3. *Research/New Insights* [Research, types of data, and data analysis tools used]

4. *Serendipity/Research Practice* [Where participants get their ideas, find new sources, role of digital tools]

5. *Role of the Library* [Use of library services and resources]

6. *Social Media*

   [General questions]

   6.1 Do you use social media?

       6.1.1   If yes, which sites and why?

       6.1.2   If no, why not?

       6.1.3   What social media do you use most frequently to communicate with other DH?

   [Twitter]

   6.2 You follow _________ (#)

   6.3 And are followed by _______ (#)

   6.4 Do you invest a lot of time in Twitter?

   6.5 How often do you tweet?

   6.6 Do you use Twitter for

       6.6.1 Clarifying own ideas?

       6.6.2 Dissemination of your own research?

       6.6.3 Sharing other's research?

6.6.4 Do you find information useful to your research?

6.6.5 Do you find information that you were otherwise unaware of?

6.6.6 Can you recall an example? (critical incident)

6.6.7 Connecting people within DH?

6.6.8 Connecting people outside of DH?

6.6.9 Connecting with larger public?

6.6.10 Meeting new people?

6.6.11 Random or relevant and important connections?

6.7 Are there some things that Twitter does not work well for?

[Remaining questions relate to Facebook and blog use]